# From Misalignment to Synergy:
# Analysis of Patents from Indian Universities & Research Institutions

Shoyeb Khan*, Satyendra Kumar Sharma*, Arnab Kumar Laha*+

## Abstract

Indian Universities and Research Institutions have been the cornerstone of human resource development in the country, nurturing bright minds and shaping the leaders of tomorrow. Their unwavering commitment to excellence in education and research has not only empowered individuals but has also made significant contributions to the overall growth and progress of the nation. Despite the significant strides made by Indian universities and research institutions, the country still lags behind many developed nations in terms of the number of patents filed as well as in the commercialization of the granted patents. With 34 percent[1] of students choosing STEM fields in India, and over 750 Universities and nearly 40,000 colleges, the concentration of patent applications in only a few top 10 institutions raises concerns.

Innovation and technological advancement have become key drivers of economic growth and development in modern times. Therefore, our study aims to unravel the patent landscape of Indian Universities and Research Institutions, examining it through the lens of supply and demand for innovations and ideas. Delving into the dynamics of patent filing and innovation trends, this study seeks to shed light on the current state of intellectual property generation in the country's academic and research ecosystem.

## 1 Introduction

Innovation has been a driving force behind advancements in society. From the invention of fire and the wheel in prehistoric times to the development of electricity, the internet, and beyond, innovation has shaped the way we live, work, and interact with the world. Many firms focus on innovation to create and sustain a competitive advantage and achieve continuous growth[1]. Patenting activity encourages and incentivizes innovation by providing a means for inventors or organizations to protect their innovations and gain a competitive advantage in the marketplace. It provides a legal framework for inventors to commercialize their inventions, recoup their investment in research and development, and potentially generate revenue through licensing or selling their patented technology. However, the growing complexity of modern technology, which often involves multiple technologies, has made it difficult for firms to rely solely on their internal research and development (R&D) departments. To address this complexity, firms may need to open up their innovation process and collaborate with external partners to access diverse expertise and resources[2].

*Birla Institute Of Technology And Science,Pilani.

*+Indian Institute of Management,Ahmedabad.

1 : https://www.weforum.org/agenda/2023/03/which-countries-students-are-getting-most-involved-in-stem/

The popular choice for these external partners has been the Universities and Research institutions(URIs). Over the past decade, there has been a remarkable surge in research collaborations between companies and universities[3]. As companies have scaled back their investment in early-stage research, they have increasingly turned to universities to fill that gap, seeking to tap into the expertise of top scientific and engineering minds in specialized domains. This trend has led to a burgeoning partnership between academia and industry, with universities playing a pivotal role in driving innovation and providing cutting-edge research capabilities to address industry challenges. While this trend has been evident in many advanced and developed countries, it is unfortunate to note that developing countries like India have not experienced the same level of collaboration between companies and universities in driving innovation and research. There could be several reasons such as limited funding[4], Industry-academia gap[5], policy challenges[6], the heavy involvement of Indian IT firms in software outsourcing[7] etc. Also, there have been enough amount of study being conducted both by government-sponsored agencies as well as academic institutions on the patenting landscape of Indian URIs.

Our study takes a unique approach to analyzing patents from Indian URIs by examining them through the lens of supply and demand principles. The first research question we seek to answer is: What is the proportion of patents granted that are attributed to Indian URIs? To answer this, we will examine the patent data from various sources and analyze the contribution of Indian URIs in terms of granted patents. This will provide insights into the extent to which these institutions are actively engaged in patenting their inventions and contributing to the overall patent landscape in India.

We will explore the most promising fields of inventions from Indian URIs. By analyzing the patent data, we will identify the fields or domains where these institutions are making significant contributions in terms of inventions that have the potential for commercialization and societal impact. This analysis will shed light on the areas where Indian URIs are excelling in terms of innovation and patenting.

Furthermore, we will investigate the most cited field of inventions from Indian URIs. By examining the citation data of patents, we can determine the fields or domains where the inventions from these institutions are being widely recognized and acknowledged by the global research community. This analysis will highlight the areas where Indian URIs are making notable contributions that are recognized and valued by the wider scientific and technological community.

As per Figure 1, 31% of patent applications originate from Other Academic institutions (including private institutions), while approximately 40.4% come from prestigious institutions like IITs/NITs, CSIR, and other government agencies. This finding is significant in the context of market pull or demand generation for innovation or patents. We argue that market pull or demand for patents or innovations can be effectively generated through sufficient government policies. Indian URIs, such as IITs/NITs, are public research institutions that advocate for knowledge accessibility to all and express concerns that patents may hinder access to new technologies or inventions. Thus, if demand for patents is stimulated by the government, which represents the public, through its policies, it may encourage patenting activities by these institutions.

Figure 1: Patent Application Distribution among Indian URIs

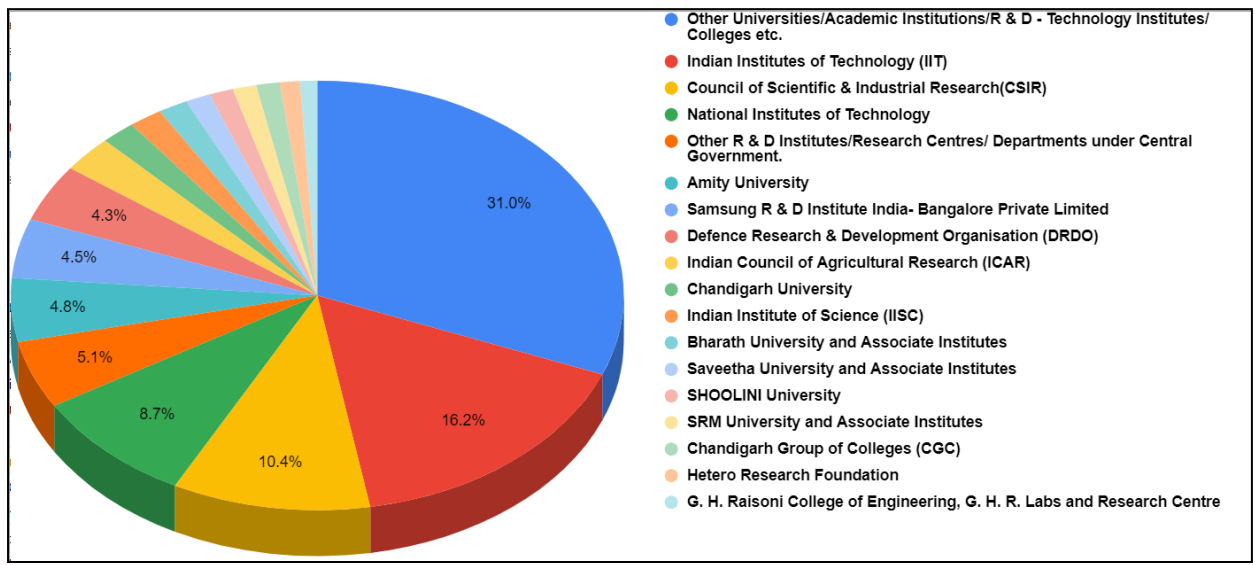

Data Sources: Open Government Data (OGD) Platform India.

Therefore, we will also investigate the various government policies or strategies in place for fostering innovations, ideas, and patents in India. We will review the existing policies and initiatives undertaken by the Indian government to promote innovation, provide support for patent filing, and encourage collaboration between academia, industry, and research institutions. This analysis will provide an overview of the policy landscape in India and its impact on the generation and commercialization of patents.

Next, we will explore the key priority technologies identified by the Indian government, where both Indian industries and Higher Education Institutions (HEIs) should focus their efforts. This analysis will highlight the technological domains that are considered strategically important for the growth and development of the country, and where Indian universities and research institutions can align their research and patenting efforts.

**2 Background**

Table 1 provides a comprehensive summary of past and future innovation waves spanning from the 19th to the 21st century, as per the Global Innovation Index(GII) 2022 report[8]. Focusing on the 21st century, which represents the Digital Age Wave, the emergence of Information and Communication Technology (ICT) is notable. ICT unfolded in two ways during this Digital Wave. Firstly, ICT served as a research tool, significantly impacting scientific advances and R&D in fields like bioinformatics, pharma, green tech, and others, with the convergence of ICT, bio- and nanotechnology, and cognitive science research. Secondly, ICT acted as a general-purpose technology, impacting non-ICT sectors through automation and AI, large-scale factory digitalization, 3D printing, and advanced robotics[9]. This analysis is significant in exploring the representation of patents or inventions from Indian Universities in the Digital Age Wave.

Table 1: Wave of Innovation Across Various Time Periods

| 19th Century | | 20th Century | | 21st Century | |
|---|---|---|---|---|---|
| Start(19th) | Mid(19th) | Start(20th) | Mid(20th) | Start(21st) | Mid(21st) |
| Steam Engine | Railroad | Chemicals & Pharma | Electronics | Digital Age Wave (ICT installation and advanced ICT adoption) | Deep Science Wave (Major scientific breakthroughs in hard sciences) |
| Textiles | Steel | Oil | Aviation | Phase 1<br><br>Diffusion of ICT network and hardware ICT installation | Phase 1<br><br>Scientific breakthrough in bio, nano-tech, health, new materials. |
|  | Electricity | Automobiles | Mass & just-in-time production | Phase 2<br><br>Adoption of advanced ICT solutions, e.g., AI, digital transformation | Phase 2<br><br>Applications of breakthroughs in health, agri-food, clean tech, transport and others |
|  |  | Nuclear Power |  |  |  |

Furthermore, apart from the resurging Digital Age wave, there is a distinct possibility of a forthcoming innovation wave referred to as the Deep Science wave. This wave is anticipated to revolve around pioneering inventions and innovations in domains such as life sciences and health, agri-food, energy and clean tech, and transport. As such, we have compiled a summary of the most promising technologies that are expected to be impacted by the Deep Science wave as shown in Table 2. This consolidation is based on insights from the GII 2022 report. Some of the notable breakthroughs include: Advancements in genetics and stem cell research, nanotechnology, biologics, and brain research are opening up new possibilities for disease detection, prevention, and cure, including the development of vaccines[10]. Novel materials, such as resins and ceramics, are being developed at the nanotechnology level, drawing on advancements in graphene and material sciences[11]. There is an unprecedented convergence of fields such as biology, agronomy, plant science, digitalization, and robotics, which is transforming the landscape of innovation in the agriculture and food industry[12].

Table 2: Promising new technologies identified by sector in which Deep Science Wave can impact.

| Technology | Digital Age Wave Impacts | Deep Science Wave Impacts |
| --- | --- | --- |
| Information and communication | Originating Sector | Use of nanotechnology and neural networks. |
| Agriculture | Automation and big data to make better decisions. | New-generation sequencing. Bioreactor-based synthetic food production. Lab-grown real meat and other future foods with higher yields and better nutrient content. Self-fertilizing crops. Precision farming. Smart fertilizers. |
| Manufacturing | Automation, Advanced Robotics and 3-D Printing. | Nanotech, new materials etc. |
| Wholesale and retail | E-commerce and Digital supply chains and logistics. | Uncertain |
| Finance and Insurance | FinTech, Digital Currencies, Block Chain. | Uncertain |
| Government | e-Government. | Uncertain |

According to the findings of the Lexi Nexis Innovation Momentum 2022 report[13], our analysis of the distribution of technologies across different geographical areas revealed an uneven distribution, with certain regions focusing on specific technological domains. Notably, the pharmaceuticals industry emerged as the largest sector with 17 innovators, primarily from the Americas, none from Asia, and four from EMEA. These observations are depicted in Figure 2.

The strong focus on electronics in the Asian region can be attributed to various factors such as established electronic manufacturing hubs, a large consumer base for electronic products, and a growing demand for technology and innovation in the region. As for its impact on India's innovation or patent landscape, it is possible that the focus on electronics in Asia could have influenced India's innovation and patent trends in this domain. India, being a major player in the global electronics industry, has witnessed significant growth in the development of electronic technologies, products, and innovations. This is evident in the increasing number of patents filed by Indian entities in the field of electronics(which we have investigated in the later section), as well as the rise of electronics-related research and development activities in academic institutions and research organizations across India.

However, it is important to conduct further research and analysis to establish a direct correlation between the focus on electronics in Asia and its impact on India's innovation or patent landscape. Factors such as government policies, economic conditions, technological capabilities, and market demand also play significant roles in shaping the innovation and patent landscape in India.

Figure 2: Industry Sector Distribution of Top 100 Companies.

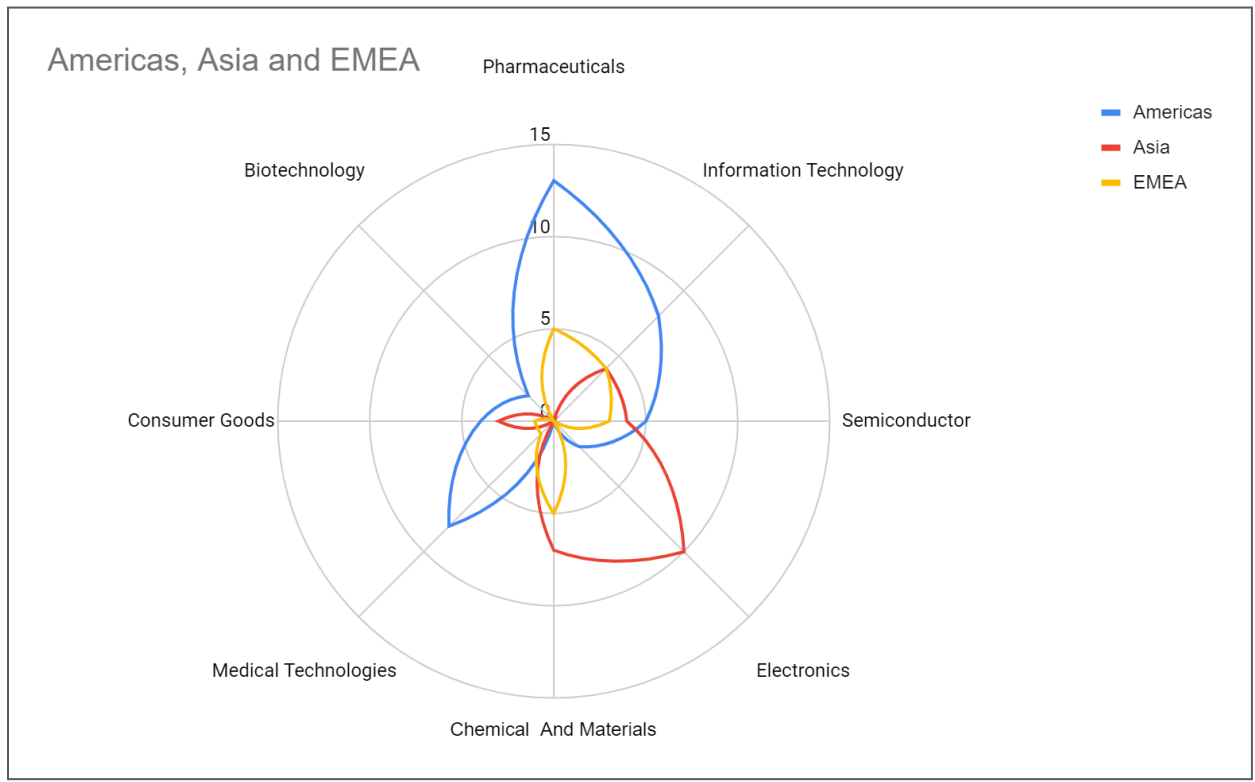

## 3 Most Valued Fields of Invention

In order to find the most valued field of invention we referred to the dataset which was developed along with the paper[14]. Patent citations are widely recognized as a reliable indicator of the scientific value of a patent, and they are also found to be significantly correlated with future citations, providing meaningful economic insights into the measure of innovation[15]. Our research findings, which are consistent with the insights presented in the GII 2022 report, reveal that the most valued technologies in our dataset are largely from the field of ICT, aligning with the ongoing Digital Age Wave of the 21st century. This underscores the importance for Indian URIs to also shift their focus towards the upcoming Deep Science Wave, which is anticipated to bring breakthrough inventions and innovations in life sciences, health, agri-food, energy, clean tech, and transport. These inventions are expected to hold significant value in the near future, making it imperative for Indian URIs to actively engage in research and development in these areas to stay at the forefront of technological advancements and maximize their contributions to innovation and economic growth.

Figure 3: Most valued field of inventions.

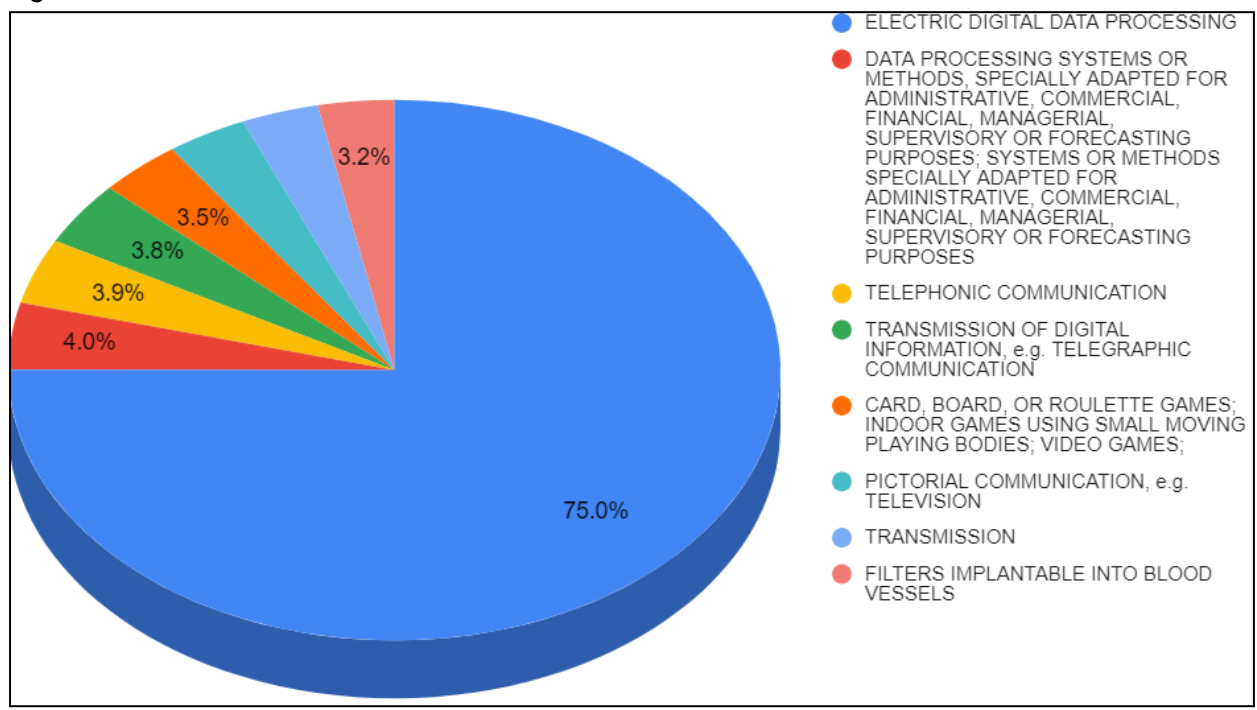

The figure depicts the most valued fields of invention till 2020. We selected the top 20 high-valued patents from the dataset available at https://github.com/KPSS2017/Technological-Innovation-Resource-Allocation-and-Growth-Extended-Data/blob/master/KPSS_2020_public.csv.zip. Further, we looked at the CPC/IPC classification code to map the field of the invention to the patent.

## 4 Indian Inventors

As per the annual report[16] published by the Office of the Controller General of Patents, Designs, Trademarks and Geographical Indications India, the following conclusions can be drawn. Firstly, the total number of patent applications filed in the fiscal year 2021-22 was 66440, which reflects a significant increase of 13.57% when compared to the filing figure of 58503 in the previous fiscal year of 2020-21. Notably, patent filings have shown moderate to substantial growth across various field of inventions, with Computer Science & Electronics, Communication, Mechanical, and Electrical fields exhibiting particularly strong growth during the year. Out of the total 66440 patent applications filed in the reported period, the number of applications filed by Indian applicants was 29508, which reflects a significant increase of 21.3% compared to the figure of 24326 in the previous year. Domestic filings by Indian applicants accounted for 44.41% of the total applications, up from 41.58% in the previous year. This consistent upward trend in Indian applicants' filings demonstrates a remarkable increase. On the other hand, the number of patent applications filed by foreign applicants during the same period was 36932, showing a growth of 8.06% compared to the figure of 34177 filed in the fiscal year 2020-21.

Secondly, Out of the total number of ordinary applications filed by Indian applicants during 2021-22, Tamil Nadu occupies the first position while Maharashtra and Uttar Pradesh occupy second and third place, respectively. This year states like Tamil Nadu showed a remarkable leap in filing compared to last year and occupies first place in the list. Similarly states like Karnataka, Punjab, Telangana and Haryana along with UTs like Chandigarh, Jammu & Kashmir and Daman & Diu also increased their filing, hence contributing immensely to the overall patent applications filed by Indian applicants.

Figure 4: Distribution of Field of Inventions by the Indian Inventors.

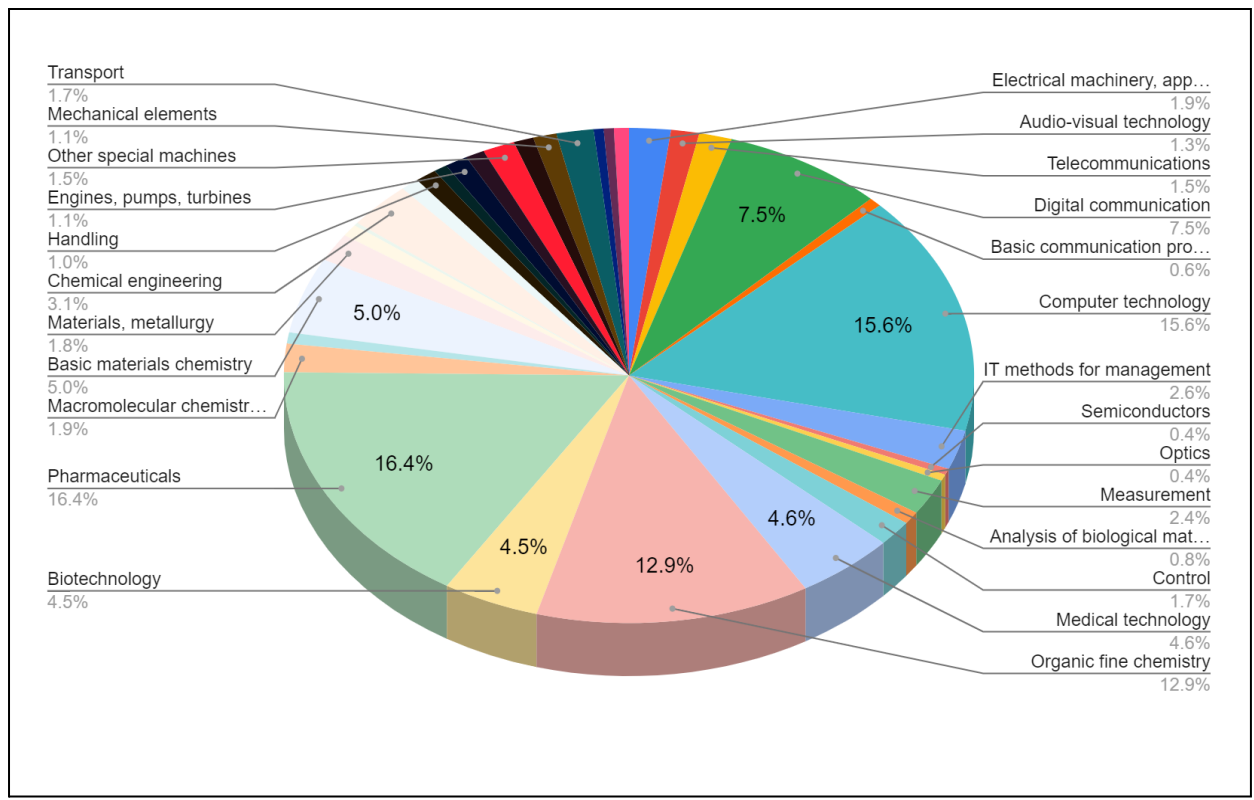

The figure depicts the field of innovation as per the patent filed by the Indian inventors. We compiled the dataset from various sources such as USPTO, Google Patents, WIPO, etc. Further, we looked at the Cooperative Patent Classification(CPC)/International Patent Classification(IPC) codes of these patents to map the field of the invention to the patent. This was done to find the answer to this question: what are the most popular choices of field of invention for Indian inventors?

The findings from Figure 4 reveal that Pharmaceuticals, Organic Fine Chemistry, Computer Technology, and Digital Communication are popular fields of invention among Indian inventors. However, there is also a noticeable diversity in the choice of fields for patent filing, with a significant increase in the number of patents granted after 2016, as shown in Figure 5. While there could be multiple reasons for this trend, one of the reasons that we found interesting and relevant to our study was the National Intellectual Property Rights Policy(NIPRP) 2016 [17]. This policy aims to promote IPR awareness, generate IPRs, establish a legal and legislative framework, improve administration and management, facilitate commercialization of IPRs, enhance enforcement and adjudication and foster human capital development.

Figure 5: Patents Granted to Indian Inventors.

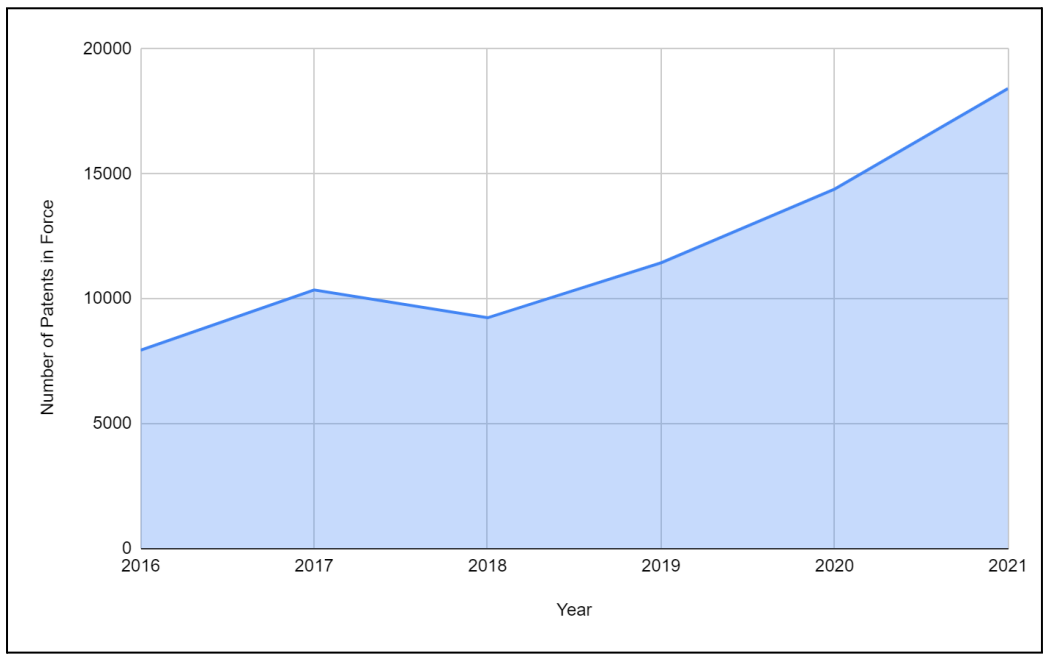

The source of data for the above figure is WIPO Statistics for India.

In addition to our analysis of the popular choice of fields of innovation, we sought to further investigate this area in greater detail. To achieve this, we examined the choice of fields of invention by Indian inventors over the years using Figure 6. This analysis allowed us to gain insights into the consistent preferences of Indian inventors when it comes to fields of innovation. Our findings indicate that fields such as biotechnology, chemistry, communication, computer technology, electrical engineering, electronics, mechanical engineering, and metallurgy have been consistently popular choices for innovation among Indian inventors.

The NIPRP 2016 includes several positive measures, and one particularly relevant to our study is the incorporation of IP creation as a crucial performance metric for public-funded R&D entities and technology institutions. The policy proposes a gradual extension of this evaluation to the patent applications filed from Tier-1 & Tier-2 institutions, reflecting a strategic focus on incentivizing and promoting IP generation in research and development activities. This step highlights the policy's emphasis on fostering a culture of innovation, intellectual property protection, and commercialization in academic and research institutions.

Figure 6: Field of Inventions of patent applications by the Indian Inventors.

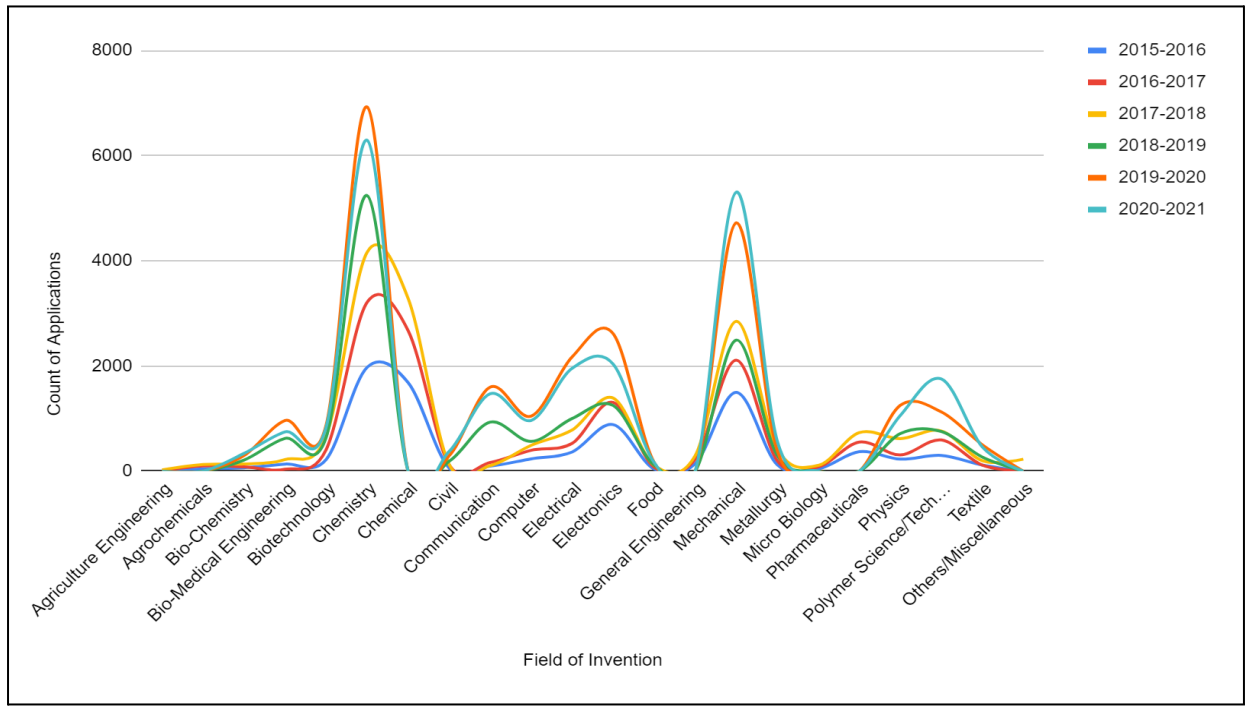

Source of dataset: Lok Sabha Unstarred Question No. 4623, dated 24.03.2021.

Figure 7: Field of Inventions of patents granted by the Indian Inventors.

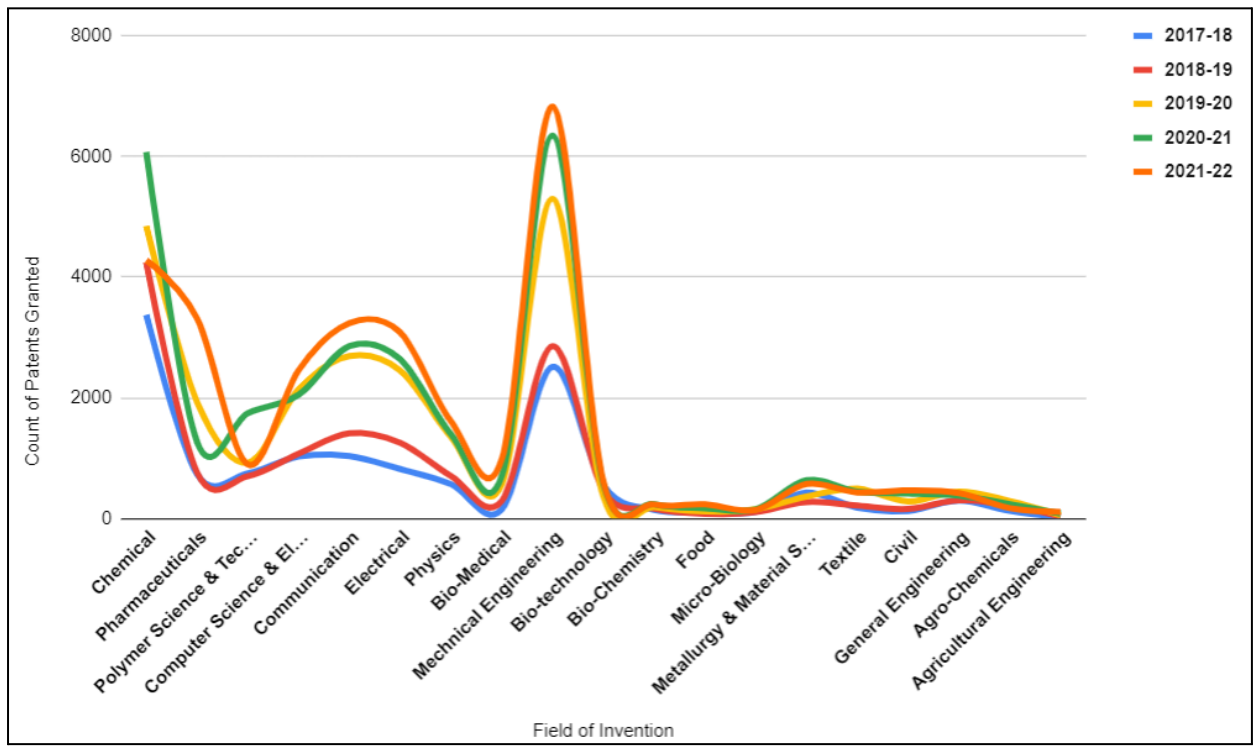

Source of dataset : [Annual Report 2021-22 by Indian Patent office.]

Figure 8 presents a contrasting perspective, as it reveals that despite the NIPRP 2016 being successful in generating IPRs from Indian inventors, the contribution of Indian University Research Institutions (URIs) to this success has been relatively low, accounting for only approx. 7% of the total patents granted to Indian inventors over a period of five years (2016-2021). Moreover, this contribution has remained relatively steady without showing any significant upward trend. Although our study is purely analytical in nature, we will strive to find data-supported answers to the questions raised in the introduction section of this paper. Through our investigation, we aim to determine whether the patent landscape at Indian URIs is aligned or misaligned with critical government policies, thereby shedding light on the possible reasons for this disparity.

In our study, we will focus exclusively on TIER-1 Indian academic institutions due to the availability of data and the concentration of granted patents to these institutions. Additionally, the evaluation of the IP landscape for these institutions is provided in the NIRF ranking report. However, we acknowledge the importance of future research examining the patenting landscape at TIER-2 institutions, and we strongly encourage such studies to further broaden the understanding of IP generation and innovation in the Indian academic ecosystem.

Figure 8: Academic contribution to granted patents.

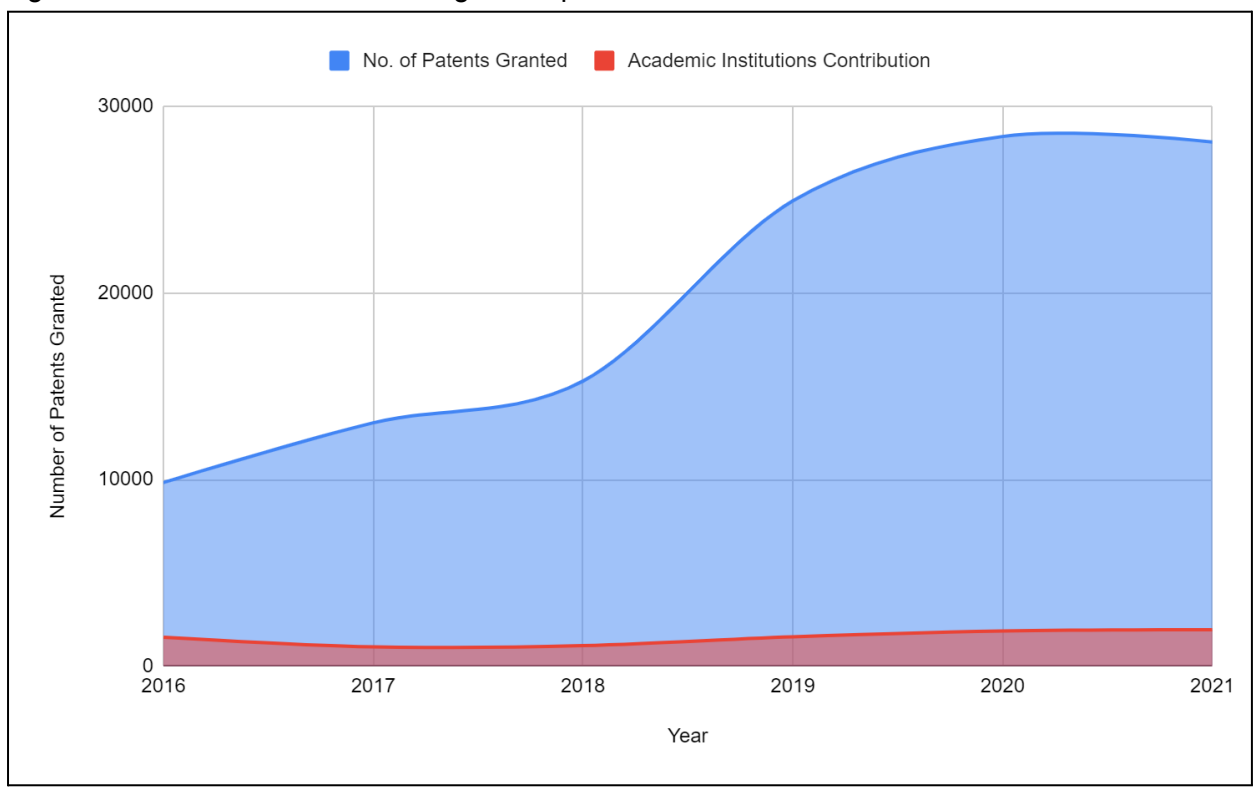

Source of dataset: WIPO Statistics for India.

## 5 TIER-1 Indian Academic Institutions

We looked at the top 10 Indian URIs as reported by NIRF 2022 and the marks allocated to them in the IP category.

We also extracted the data from the WIPO about the applications submitted by these institutions both at the domestic patent offices and international patent offices. As per Figure 9, we find that IIT Madras, IIT Bombay and IIT Delhi have shown the most promising results to the NIRF committee. As per Figure 10, we find that the choice of patent office for patent application has been the home country office i.e. Indian Patent Office. Also, it is worth noting that IIT Bombay has submitted approx. 1092 domestic patent applications during the period 2016-2023. When examining the field of inventions, our analysis reveals that ICT, Healthcare, Mechanical Engineering, and Energy & Environment have been popular choices for innovation among these institutions, as depicted in Figure 11. This provides insights into the most promising field of inventions from Indian universities and research institutions, addressing the question of our study.

Figure 9: IPR: Patents Published & Granted Marks Allocated to various Institutions.

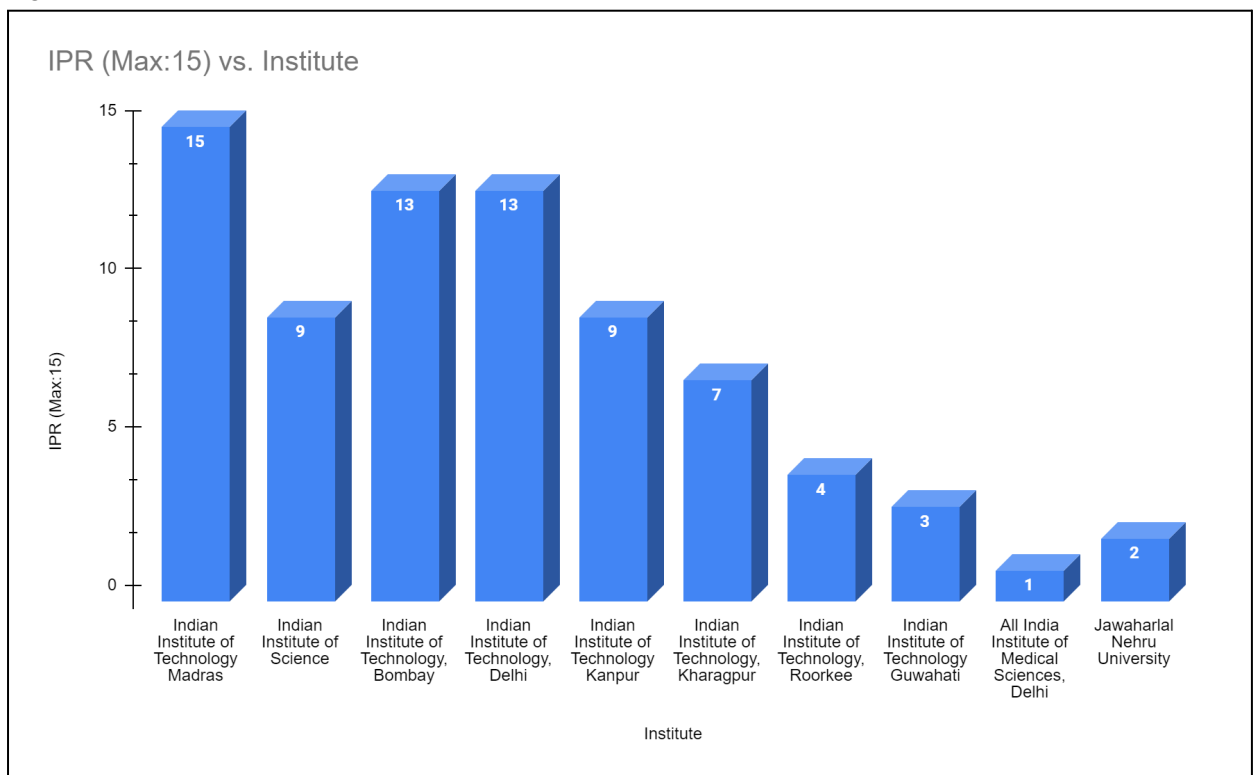

The figure depicts the marks allocated to the respective institutes in the IPR category by the NIRF committee. The order of these institutions is as per the overall ranking provided by the NIRF ranking report 2022.

Figure 10: Patent Applications Filed by Indian URIs (2016-23).

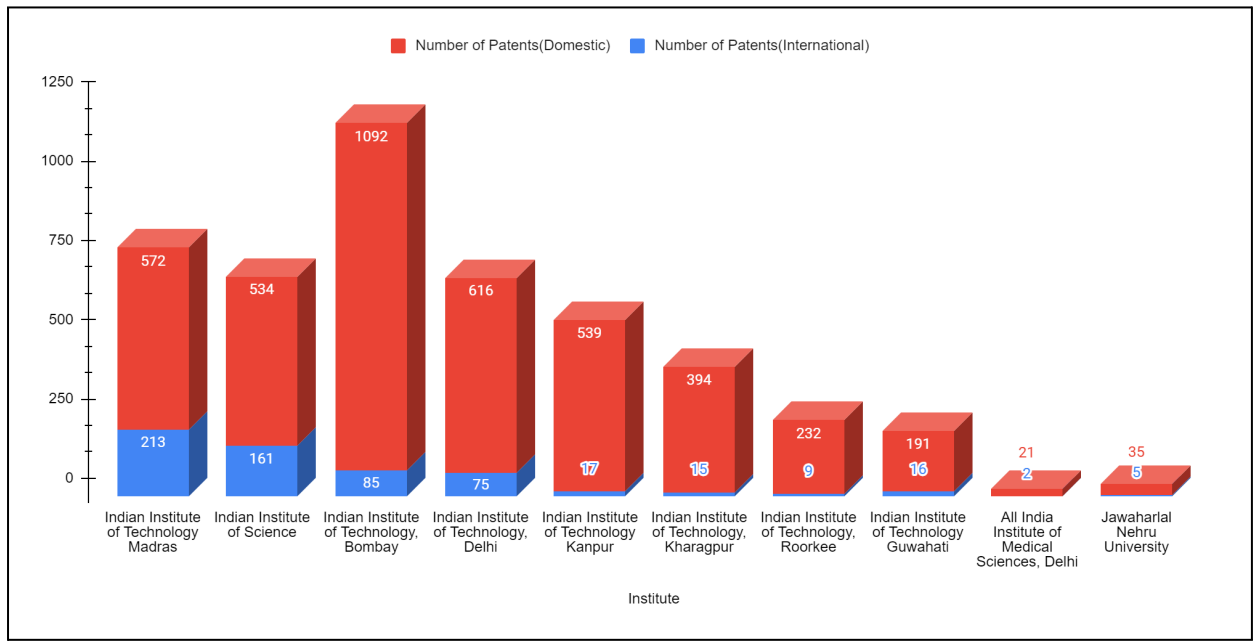

The figure depicts the proportions of the patent applications filed at the Indian Patent Office (Domestic) to the patent applications(aggregate) filed at various foreign patent offices. Source of the dataset: WIPO Statistics 2022.

Figure 11: Promising Field of Inventions (2016-23).

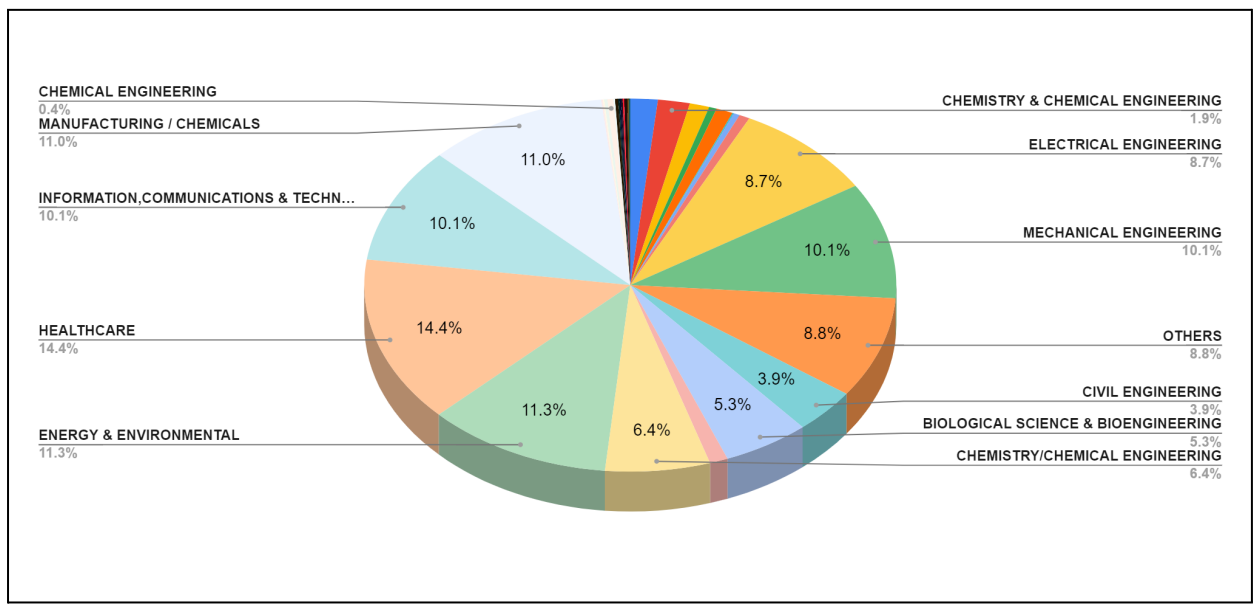

The figure depicts the popular choices of field of invention by the Indian Universities(Top 10 institutions as per NIRF ranking report 2022). We looked at the CPC/IPC codes of the patent applications and also would like to mention the figure doesn't represent the CPC/IPC codes of the granted patents. Source of Dataset: Various sources such as WIPO, Google Patents, Respective TTO websites etc.

Figure 12: Most Cited Field of Inventions (2016-23)

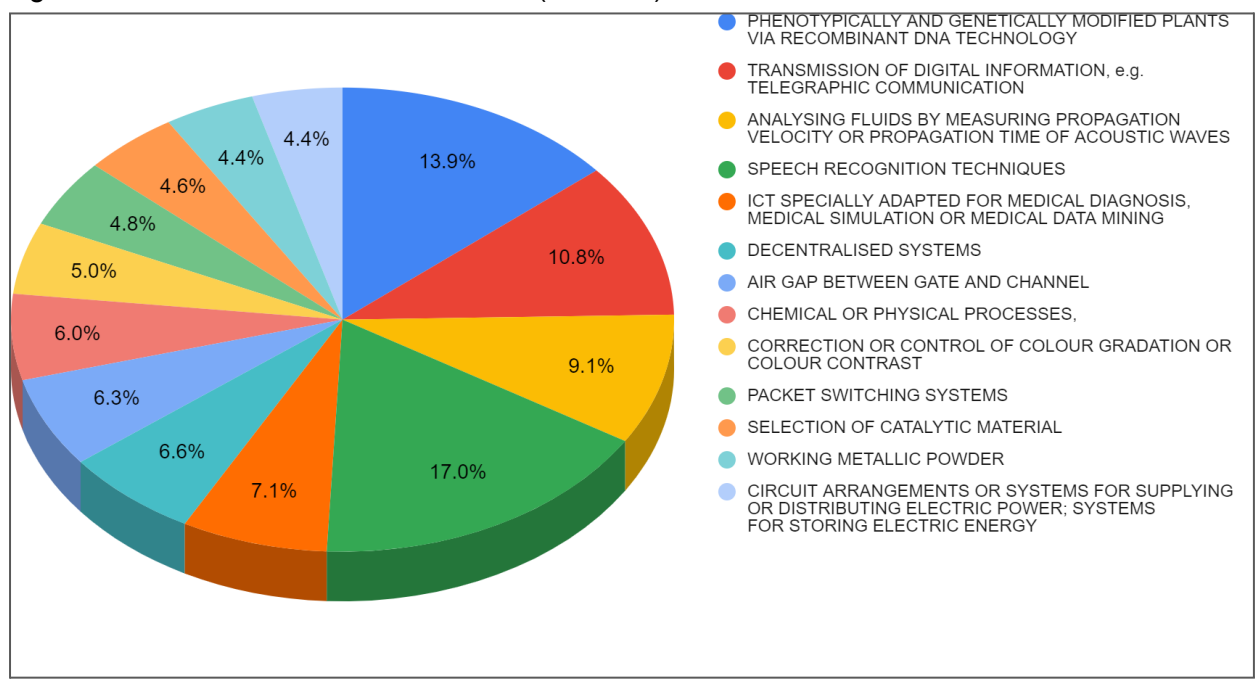

The figure depicts the field of inventions which belongs to the Indian URIs patents granted over the period of 2016-23. We referred to the Google patents platform and extracted the 'cited by' count of the patents granted to the Indian URIs.

Earlier we have mentioned that the future citation can act as the indicator for innovation measures or patent value. Therefore, we analysed the citation received by the granted patent of Indian URIs. We analysed the granted patents of Indian URIs data on the Google Patents platform. One nice feature of this platform is that it provides information on the citations received by the patents. Also, we firmly believe that if a patent is receiving a lot of citations implies it is receiving a lot of attention and hence the market value of the patent will be more as compared to other patents. But we also take cognizance of the fact that patent valuation is a kind of 'black art' i.e. a lot of parameters are considered when it comes to valuing a patent and different agencies have adopted a different method for the same. However, for academic research 'cited by' feature can serve as a good indicator for analysis.

Figure 12 depicts some of the most cited field of inventions coming from the Indian URIs. We could see a glimpse of the Deep science wave age with Recombinant DNA technology receiving the most citations by other inventors in the same field. Since we are in the Digital age wave, there was no doubt that we could see the ICT field of invention in Figure 12. Also, Speech Recognition technologies are most popular among inventors from other regions. With this, we are able to understand the most cited field of inventions from Indian URIs.

Based on the analysis of Figure 13 and Figure 14, we found that IIT Madras has shown a preference for the field of inventions such as material science, electrical communication systems, and data processing. This finding holds significance in light of the ICT policy of 2018 in the state of Tamil Nadu where the institute is located. We also observed that sponsored research is the most common route for patent development at Indian URIs, and many of these sponsored research initiatives are driven by government policies. In a similar vein, our analysis revealed that IIT Bombay has shown a preference for the field of inventions such as medical science, semiconductor technology, and chemical sciences. This finding is noteworthy in the context of the Electronics Policy of 2016 and IT & ITES Policy of 2015 in the state of Maharashtra, where the institute is located. These policies may have influenced the choice of field of inventions at IIT Bombay, indicating the role of regional policies in shaping patent trends at academic institutions.

Thus, we would like to highlight the importance of regional policies in shaping the field of inventions at academic institutions. It is crucial for such policies to be tailored according to the local infrastructure and facilities available for conducting sponsored research on a large scale. By creating a unique ecosystem that supports institutions in the region and identifies niche technological expertise, regional policies can foster innovation and promote regional technological growth. Fine-tuning regional policies based on the specific needs and strengths of the region can optimize the impact of sponsored research and contribute to the overall development of the technological landscape.

Figure 13: Field of Innovations available for commercialization by IIT Madras.

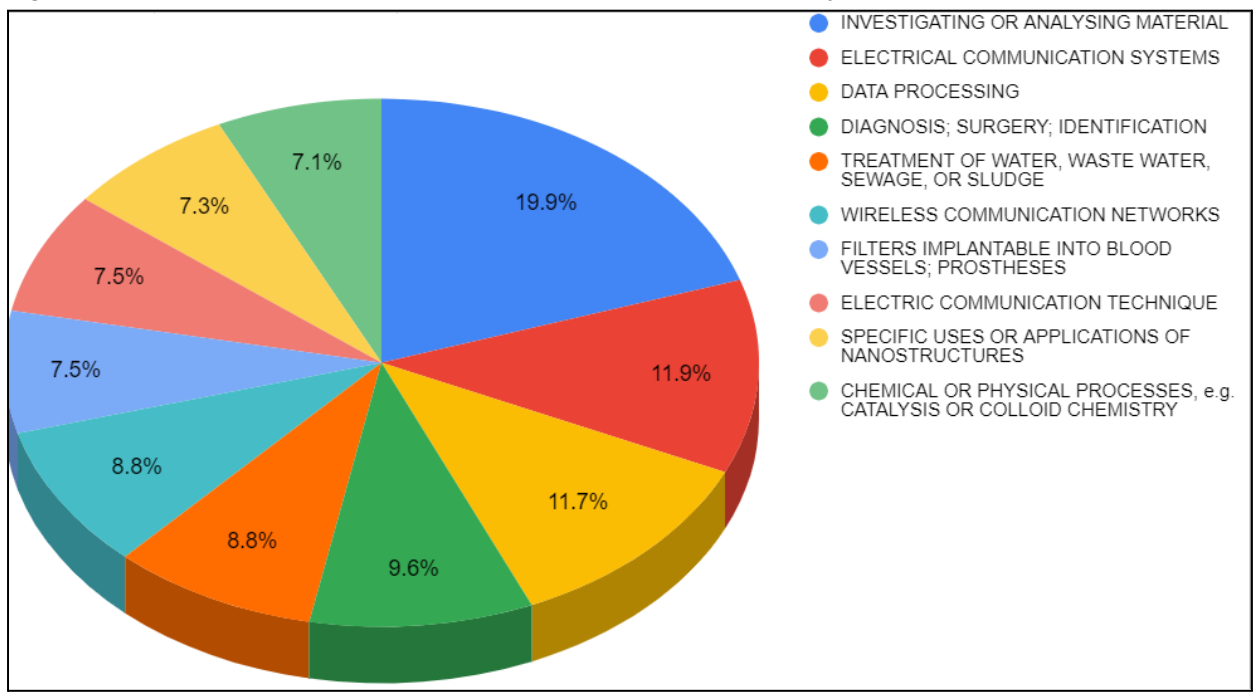

Source of Data: IIT M TTO website. (Available at: https://icandsr.iitm.ac.in)

Figure 14: Field of Innovations available for commercialization by IIT Bombay.

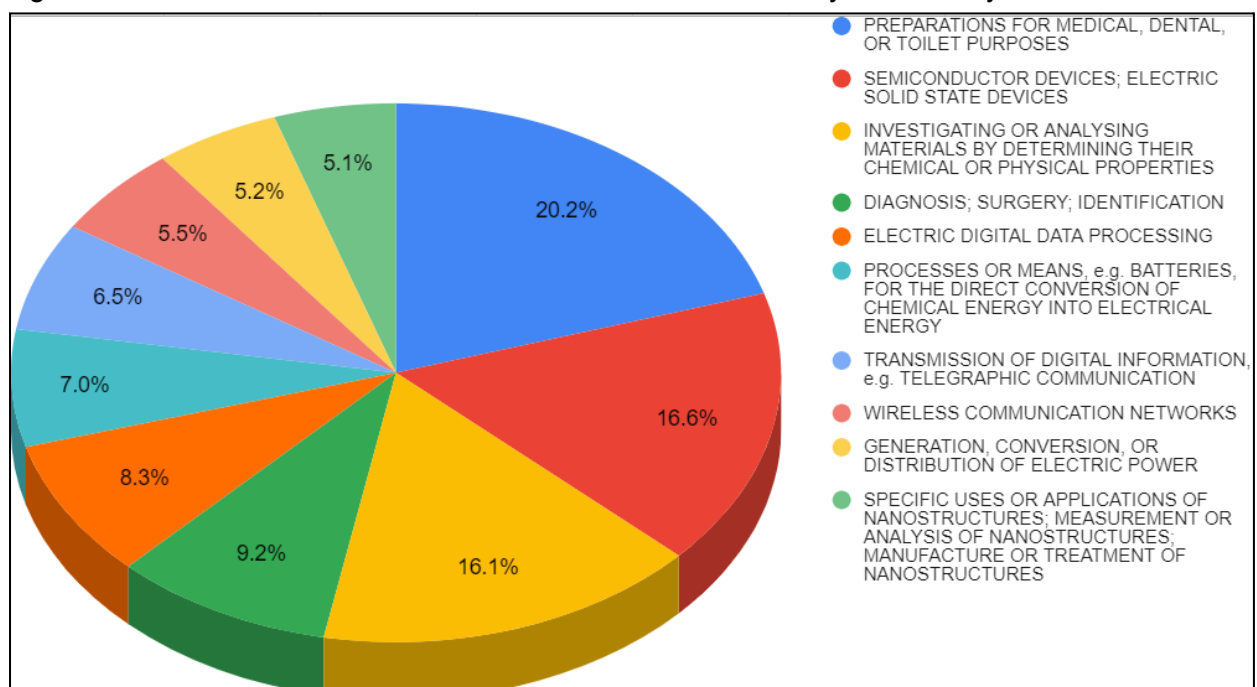

Source: IIT Bombay TTO Website. Available at :
(https://www.ircc.iitb.ac.in/IRCC-Webpage/rnd/LicensingTechnologies.jsp)

**6 National Strategy for Most Promising Technologies**

Table 3: Priority Technologies for the Indian Government.

| Technology | Policy |
| --- | --- |
| Artificial Intelligence | National Strategy for Artificial Intelligence 2018 |
| Blockchain | National Strategy on Blockchain 2021 |
| Biotechnology | National Biotechnology Development Strategy 2021-25 |
| Quantum Computing | Evolving National Strategy |
| 5G Networks | Part of Telecommunications Policy |
| Nanotechnology | Part of Nanotechnology Initiatives Policy |
| Renewable Energy | Part of Energy Policy |
| Cyber Security | Evolving National Strategy |
| Robotics | Part of National Strategy for Artificial Intelligence |

The Science, Technology, and Innovation Policy (STIP) 2013 had a specific focus on critical R&D areas such as agriculture, telecommunications, energy, water management, health and drug discovery, materials, environment, and climate variability. However, with the introduction of STIP 2020, the focus has been divided into two areas. The first area is strategic technologies, including space technology, nuclear technology, biotechnology, and cyber-physical domains. The second area is disruptive technologies, including blockchain, AI, 3D printing, quantum computing, and IoT. STIP 2020 emphasizes that the country cannot solely rely on importing strategic technologies from abroad indefinitely. Instead, it emphasizes the need to effectively utilize the expertise of academia and industry in the strategic sector. The policy also highlights the fact that Indian universities generally do not pursue R&D beyond Technology Readiness Level(TRL)-1, creating a significant gap in the domain of commercialization and understanding of TRL among scientists, researchers, and developers. The languages, risks, and scales of funding vary at different levels, including university level, industry level, and grassroots, which further adds to the challenges in this domain.

The aim of this study is to find the shift in the priorities in the technological landscape and whether the Indian URIs patenting landscape is aligned with this shift or not. To highlight this shift, we will specifically examine three key strategies implemented by the GOI in the context of disruptive technologies.

## 6.1 National Strategy for Artificial Intelligence

As per the report[18] published by the NITI Aayog, In order to boost both core and applied research in Artificial Intelligence (AI), a two-tier integrated approach is proposed. The first tier consists of Centres of Research Excellence in AI (COREs), which will focus on core research and source fundamental knowledge/technologies to keep India prepared for future technologies. The second tier consists of the International Centre for Transformational Artificial Intelligence (ICTAI), which will focus on application-based technology development and deployment, including the commercialization of ideas/concepts or prototypes. Additionally, an umbrella organization called the Centre for Studies on Technological Sustainability (CSTS) could be established to address issues related to finance, social sustainability, and global competitiveness of the technologies developed, similar to programs like CREATE in Singapore or Innovate UK.

The COREs for AI will initially focus on core research in evolving and new areas of AI, with proposed establishments at IISc, ISI, top IITs, and IIITs. These COREs should also establish linkages with premier institutions in other disciplines, such as AIIMS for healthcare, TISS for arts and social sciences, etc., as AI research requires a multidisciplinary approach. Furthermore, these COREs should also act as guides and mentors for other institutes conducting AI research, following a hub-and-spoke model, to facilitate the broad-based development of AI research capabilities across India. Possible focus areas for the COREs for AI could include Sensory AI (Computer Vision, IoT etc.), Physical AI (Robotics, Industrial Automation etc.), Cognitive AI (NLP, worker training etc.), General AI, High precision learning from small data sets, Research on new algorithms (e.g. advance cryptography, security), data sets etc., and Explainable AI.

The proposed National AI Marketplace (NAIM) is suggested to consist of three distinct modules in its early stages of development: a) Data marketplace, b) Data annotation marketplace, and c) Deployable model marketplace / Solutions marketplace. These modular marketplaces will support the AI development value chain by facilitating collaboration, reducing the time and cost of data collection and annotation, and providing a platform for multiple solutions deployment for scale and network effect. The goal of NAIM is to ease adoption efforts for all participants, including private enterprises, PSUs, governments, startups, and academia, by providing a common platform for collaboration, solution building, and adoption at scale.

Furthermore, the proposed marketplace can also benefit academic researchers significantly, as they will gain valuable insights into industry challenges and access to real-world industry data, enhancing their research capabilities. The standardized and ubiquitous problems that most organizations face in their activity streams, such as object detection in images or video streams, conversational smart chatbots (text and speech), speech-to-text and text-to-speech, assistive diagnostic solutions, language recognition and transcription, contextual data mining to discover complex patterns, price optimization, data collection, curation and annotation for specific business use, supply and demand forecasting, and server, app and web uptime/downtime prediction, can be addressed through the NAIM marketplace.

### 6.2 National Strategy on Blockchain

As per the report [19] published by the Ministry of Electronics & Information Technology(MeitY), the development of a National Blockchain Framework (NBF) is proposed as a collaborative effort involving various stakeholders, including the Government, premier academic and research institutions, startups, and industry. Startups and innovators will be encouraged to collaborate through grand challenges to promote innovation and use case development in various domains. The Ministry of Electronics and Information Technology (MeitY) will initiate research projects focusing on advanced research in Blockchain technology, addressing challenges such as standards and interoperability, scalability and performance, consensus mechanisms, security and privacy, and detection of vulnerabilities, through its working group mechanism. The need for an NBF is to aid in scaling up deployments, developing shared infrastructure, and enabling cross-domain application development, initially for government applications and later extending to other relevant applications. The proposed infrastructure will host multiple Blockchain platforms, including an indigenous blockchain platform with best practices and advanced technological features, domain-specific chains controlled using smart contract logic, and generic value-added features such as proof-of-storage, proof-of-existence, and predictive visual analytics and intelligence, along with integration points with existing national services and security/privacy measures.

## 6.3 National Biotechnology Development Strategy

As per the reports [20] published by the Department of Biotechnology(DBT), Ministry Of Science And Technology(MS&T), Government of India(GOI), the strategy focuses on strengthening and nurturing a research and innovation ecosystem that encompasses various sectors, including research institutes, laboratories, startups, small and large industries, and outreach to tier 2 and tier 3 cities. Biotechnology is recognized as a key driver for a knowledge-based economy, with its potential applications in healthcare, agriculture, food, informatics, nano and forensics, among others. The strategy aims to leverage biotechnology to create favourable conditions for sustainable development and deployment of biotechnologies, with a specific focus on the Indian perspective.

Building upon the National Biotechnology Development Strategy-I (NBDS) 2015-2020, the new strategy aims to make a quantum jump in addressing priority areas such as agriculture, food and nutritional security, health and wellness, environmental safety, clean energy, and bio-manufacturing, aligning with India's goal of becoming a USD 5 trillion economy. Special missions called Atal Jai Anusandhan Biotech (UNaTI) Missions have been launched to address significant national and global challenges related to maternal and child health, antimicrobial resistance, vaccines for infectious diseases, food and nutrition, and clean technologies, in alignment with national and global priorities.

To accelerate the translation of biotechnology research from lab to market, the strategy includes a clustering approach with flexible funding mechanisms. This involves identifying national priorities for focused biotechnology missions, directing funding towards new areas of biology such as synthetic biology, quantum biology, nutrigenomics, personalized medicine, and microbiome technologies, and bridging gaps in the progression of new chemical entities (NCEs) and phytopharma/new biologicals. The overall goal is to promote innovation, research, and development in the field of biotechnology, driving economic growth and creating a sustainable knowledge-based bio-economy.

**Conclusion**

The vision of STIP 2013 aimed at establishing a robust and sustainable Science, Research, and Innovation System for a high-technology-led path for India (SRISHTI). However, STIP 2020 has set a new vision to attain technological self-reliance and position India as one of the top three scientific superpowers in the next decade. This updated vision reflects a shift towards a more ambitious and proactive approach in positioning India as a global leader in science, technology, and innovation, with a focus on achieving self-sufficiency and excellence in cutting-edge technologies. To attain self-sufficiency, it is crucial to have ownership of technological inventions and patents that grant us this ownership. Furthermore, it is noteworthy to mention that the current alignment of Indian URIs with priority technologies is not optimal, considering the significant shift in the technology landscape, particularly with the emergence of the Deep Science Wave. However, with proactive support from government policies, we hold a strong belief that Indian URIs can realign themselves towards the development and ownership of priority technologies, thereby fostering technological advancement and self-sufficiency. And, we are of the view that regional policies, derived from national strategic policies and tailored to leverage niche regional technological expertise, can play a pivotal role in fostering innovation and driving technological growth at the regional level, with support from Indian URIs in the respective regions.

**Acknowledgement**

We would like to thank Dr Frank Tietze, Professor of Innovation Engineering, IfM, University of Cambridge for his guidance and advice related to this study.